# EXACT SOLUTION FOR ONE DIMENSIONAL MULTIBARRIER TUNNELING

Siddhant Das

siddhantdas@yahoo.co.in

Electronics and Communication Engineering, National Institute of Technology, Tiruchirappalli, TN, India.

**ABSTRACT** Quantum tunneling across multiple barriers as yet is an unsolved problem for barrier numbers greater than five. The complexity of the mathematical analysis even for small number of barriers pushed it into the realms of Numerical Analysis. This work is aimed at providing a rigorously correct solution to the general N barrier problem, where N can be any positive integer. An *exact algebraic solution* has been presented, which overcomes the complexity of the WKB integrals that are traditionally employed, and matches the earlier results reported for small number of barriers. The solution has been explored to considerable depth and many startling consequences have been pointed out for 500 and 1000 barriers. These are quite revealing and open up many avenues for engineering applications and further research.

## I INTRODUCTION

Tunneling of particles across classically forbidden regions is one of the novel implications of Quantum Mechanics. It surfaced in the work of Friedrich Hund[1] in 1927 and since then has been accepted as a very general phenomenon of Nature. Quantum Tunneling has a multitude of applications, an interesting account of which is given in the Nobel Lecture of Leo Esaki.[2] In the recent times, electron tunneling has been the central theme of several models in Molecular Biology[3][4] and is at the heart of modern electronic devices.[5][6] Calculation of tunneling probabilities for a rectangular barrier is a standard illustration in undergraduate texts in Quantum Mechanics.[7][8] The popularity of this problem has led to the adoption of a uniform set of notation in almost all publications. Extensions for double and triple rectangular barriers were carried out in 1970 to investigate humped potentials in Nuclear Fission by Ray Nix and have been a subject of exhaustive study even today.[9][10] However studies for barrier numbers larger than 5 have seldom been reported.[5][11] This is perhaps attributed to the mathematical difficulties encountered in solving tunneling problems using the conventional approach which is analyzed in Section II. In this work we present a *rigorously correct solution to the N-Barrier Problem*. The multibarrier problem springs up repeatedly in the literature especially in the analysis of a finite super lattice.[5][12] In the next section an overview of the mathematical methods employed in solving barrier problems is given and the setting for the current problem is gradually developed. The notation is explained in juxtaposition. The complete solution is traced in Section III and a discussion of results is provided in Section IV. One observes strange phenomena accompanying, very large barrier numbers like $10^2$ or $10^3$ (and even more). Some concluding remarks are provided in Section V and the approach is generalized for an arbitrary multibarrier problem. Areas for further study are also noted here.

## II PROBLEM FORMULATION AND OVERVIEW

In one dimensional Tunneling problems where the potential is piecewise constant, the wave function $\psi(x)$ is obtained by solving the time independent Schrödinger Equation (TISE) in every region. The individual 'pieces' of $\psi$ have a contribution from plane wave solutions propagating in either direction and a matching is achieved by requiring that the pieces and their derivatives be equal at the discontinuities of $V(x)$. For single and double barriers this can be done with minimal algebra, but the process gets more involved for higher barrier numbers, as the number of regions grows and more boundary constraints have to be met. For a $N$ barrier problem(NBP) one gets $2N + 1$ regions and $4N$ boundary conditions(equations). In every region $\psi$ is determined up to two complex coefficients (Section III) which gives $2(2N + 1) = 4N + 2$ coefficients. These are the probability amplitudes for the forward and backward travelling wave components that make up $\psi$ in a particular region. At infinity there is no discontinuity to offer a reflection, thus the wave function in the final region has only a forward travelling component. This sets the probability amplitude for the backward propagating wave component to zero in this region. So it reduces to effectively pinning down $4N + 1$ amplitudes. The boundary condition equations are linear and one can at best expect to get $4N$ of these in terms of one of the amplitudes (provided the equations are independent). This naturally invites matrix methods,[5][13] but leads to complications, as it requires the multiplication of long sequence of matrices ($2N$ in this case), which limit the computations to small barrier numbers. Tunneling problems are also attempted using approximations (famously the WKB method) and other numerical techniques[9], even so the solutions can only be realized for small barrier numbers.

$V(x)$ for the problem at hand is a symmetric rectangular potential array (Fig.1) which is defined to be zero for $x < 0$ and $x > N\delta + (N - 1)\tau$. Each region is labeled by an integral index (roman numerals). In the subsequent discussion $n$ denotes a general *even number*(zero included) and $m$ a general *odd number*. The $n$ indexed regions correspond to wells of width $\tau$ where $V(x) = 0$ and the $m$ regions correspond to barriers of width $\delta$ and height $V_o$. For a NBP : $0 \leq n \leq 2N$ and $1 \leq m \leq 2N - 1$.

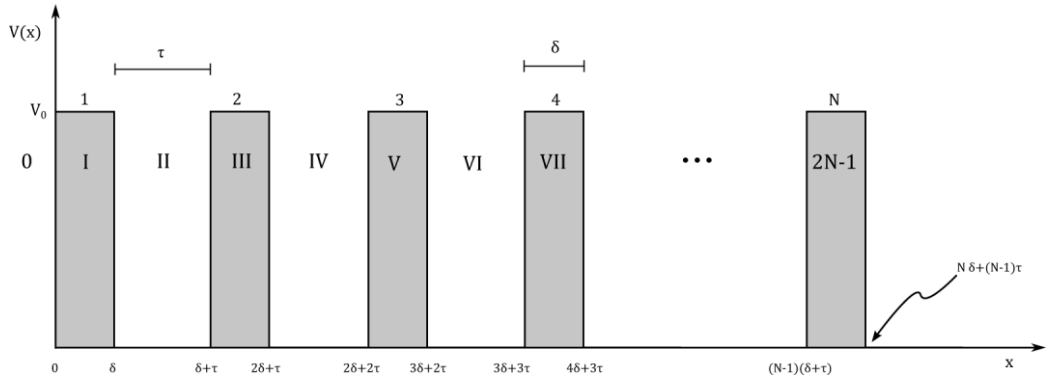

**Fig.1** $V(x)$ for a symmetric $N$ Barrier Problem(NBP). Barrier height- $V_o$. Barrier width-$\delta$. Well width-$\tau$.

### III SOLUTION

From Fig.1

$$V(x) = \begin{cases} V_o & \dfrac{m-1}{2}(\delta + \tau) < x < \dfrac{m+1}{2}\delta + \dfrac{m-1}{2}\tau \\ 0 & \dfrac{n}{2}\delta + \left(\dfrac{n}{2} - 1\right)\tau < x < \dfrac{n}{2}(\delta + \tau) \end{cases} \qquad (1)$$

The TISE in the $m$ and $n$ regions takes the form,

$$\psi'' + \gamma^2 \psi = 0 \qquad \gamma^2 = \dfrac{2\mu}{\hbar^2}(E - V_o) \qquad (2)$$

$$\psi'' + \kappa^2 \psi = 0 \qquad \kappa^2 = \dfrac{2\mu}{\hbar^2} E \qquad (3)$$

respectively, where $\mu$ is the mass of the particle. The general solutions for equations (2), (3) is written as

$$\psi_m = A_m e^{i\gamma x} + B_m e^{-i\gamma x} \qquad \frac{m-1}{2}(\delta + \tau) < x < \frac{m+1}{2}\delta + \frac{m-1}{2}\tau, \qquad (4)$$

$$\psi_n = A_n e^{i\kappa x} + B_n e^{-i\kappa x} \qquad \frac{n}{2}\delta + \left(\frac{n}{2}-1\right)\tau < x < \frac{n}{2}(\delta + \tau). \qquad (5)$$

Define ket $|V_j\rangle = \begin{pmatrix} A_j & B_j \end{pmatrix}^T$ and $|F\rangle = \begin{pmatrix} \bar{f} & f \end{pmatrix}^T$ where $f$ is a function of $x$ over a specified domain. With this notation equations (4) and (5) can be written as:

$$\psi_m = \langle \Gamma | V_m \rangle \ , \quad \Gamma(x) = e^{i\gamma x} \ , \quad \frac{m-1}{2}(\delta + \tau) < x < \frac{m+1}{2}\delta + \frac{m-1}{2}\tau, \qquad (6)$$

$$\psi_n = \langle K | V_n \rangle \ , \quad K(x) = e^{i\kappa x} \ , \quad \frac{n}{2}\delta + \left(\frac{n}{2}-1\right)\tau < x < \frac{n}{2}(\delta + \tau). \qquad (7)$$

It must be understood that the inner products are computed *point wise*. $A_j$s and $B_j$s ($j = m, n$) are complex coefficients that have to be determined by imposing boundary conditions. The boundary conditions relate $|V_j\rangle$ and $|V_{j+1}\rangle$ and can be written as

$$\begin{cases} \psi_n\left(\frac{n}{2}(\delta + \tau)\right) = \psi_{n+1}\left(\frac{n}{2}(\delta + \tau)\right) \\ \psi'_n\left(\frac{n}{2}(\delta + \tau)\right) = \psi'_{n+1}\left(\frac{n}{2}(\delta + \tau)\right), \end{cases} \qquad (8)$$

$$\begin{cases} \psi_m\left(\frac{m+1}{2}\delta + \frac{m-1}{2}\tau\right) = \psi_{m+1}\left(\frac{m+1}{2}\delta + \frac{m-1}{2}\tau\right) \\ \psi'_m\left(\frac{m+1}{2}\delta + \frac{m-1}{2}\tau\right) = \psi'_{m+1}\left(\frac{m+1}{2}\delta + \frac{m-1}{2}\tau\right). \end{cases} \qquad (9)$$

Equations (8), (9) translate into the following.

$$|V_n\rangle = \frac{1}{2\gamma} E_n |V_{n+1}\rangle \qquad (10)$$

$$|V_m\rangle = \frac{1}{2\kappa} O_m |V_{m+1}\rangle \ , \qquad (11)$$

where:

$$E_n = \begin{pmatrix} \alpha \epsilon^\beta & \beta \epsilon^\alpha \\ \beta \epsilon^{-\alpha} & \alpha \epsilon^{-\beta} \end{pmatrix} \qquad O_m = \begin{pmatrix} \alpha \eta^\beta & -\beta \eta^{-\alpha} \\ -\beta \eta^\alpha & \alpha \eta^{-\beta} \end{pmatrix}$$

$$\alpha = \kappa + \gamma \qquad \qquad \beta = \kappa - \gamma$$

$$\epsilon = e^{-\frac{n}{2}i(\delta + \tau)} \qquad \eta = e^{i\left[\frac{m+1}{2}\delta + \frac{m-1}{2}\tau\right]}$$

Alternatively one could relate $|V_j\rangle$ and $|V_{j-1}\rangle$ by evaluating (8) and (9) at $x = \frac{n}{2}\delta + \left(\frac{n}{2}-1\right)\tau$ and $x = \frac{m-1}{2}(\delta + \tau)$, respectively for $n \geq 2$. These however are *equivalent* to the above equations. Equations (10) and (11) classify the many boundary conditions into two sets that suffice to solve a second order ODE. The $m$ and $n$ dependence in the matrices is compacted in $\eta$ and $\epsilon$. Equations (10) and (11) are recognized as 'evolution' equations that carry the ket $|V_j\rangle$ to the next $|V_{j+1}\rangle$ (or the other way) via transfer matrices $E_n$ and $O_m$ (or their inverses). All transfer matrices are non-singular with a determinant (to be called $\Delta$)

$$\Delta = \alpha^2 - \beta^2 = (\kappa + \gamma)^2 - (\kappa - \gamma)^2 = 4\kappa\gamma. \qquad (12)$$

$\Delta$ is a fundamental parameter of the problem. It becomes zero at $E = V_o$.

As discussed in Section II, $|V_{2N}\rangle = \begin{pmatrix} A_{2N} \\ 0 \end{pmatrix}$ (since $B_{2N} = 0$) and $A_{2N}$ is the amplitude for the forward travelling wave component in the last region. Kets $|V_j\rangle$, $\forall j \in \mathbb{N} \cup \{0\}$ can be determined in terms of $|V_{2N}\rangle$ using equations (10) and (11) iteratively.

For instance,

$$|V_n\rangle = \left(\frac{1}{2\gamma} \boldsymbol{E}_n\right) |V_{n+1}\rangle$$

$$= \left(\frac{1}{2\gamma} \boldsymbol{E}_n\right)\left(\frac{1}{2\kappa} \boldsymbol{O}_{n+1}\right) |V_{n+2}\rangle$$

$$= \left(\frac{1}{2\gamma} \boldsymbol{E}_n\right)\left(\frac{1}{2\kappa} \boldsymbol{O}_{n+1}\right)\left(\frac{1}{2\gamma} \boldsymbol{E}_{n+2}\right)\ldots\left(\frac{1}{2\kappa} \boldsymbol{O}_{2N-1}\right) |V_{2N}\rangle . \quad (13)$$

A similar relation can be obtained for $|V_m\rangle$. It remains to obtain a compact formula for these long sequence matrix products, which is derived next.

In tunneling problems the canonical variables of interest are the transmission coefficient ($T$) and the reflection coefficient ($R$) that satisfy the identity $T + R = 1$,[14] which follows from the continuity equation for the probability current density.[15] For a NBP, $T$ and $R$ are defined as.

$$T = \left|\frac{A_{2N}}{A_o}\right|^2 \qquad R = \left|\frac{B_o}{A_o}\right|^2 . \quad (14)$$

To derive expressions for $T$ and $R$ one needs to relate $|V_o\rangle$ with $|V_{2N}\rangle$. From equation (13),

$$|V_o\rangle = \frac{1}{\Delta^N} \boldsymbol{E}_0 \boldsymbol{O}_1 \boldsymbol{E}_2 \ldots \boldsymbol{E}_{2N-2} \boldsymbol{O}_{2N-1} |V_{2N}\rangle , \quad (15)$$

setting $2\gamma 2\kappa \ldots 2\gamma 2\kappa = (4\kappa\gamma)^N = \Delta^N$ from equation (12). These operator products can be simplified by using Pauli Matrices. A compact expression(equation (18)) can be obtained by using the algebraic properties of these matrices. Most of these properties are expounded in the Feynman Lectures on Physics.[16] The three Pauli Matrices along with the $2 \times 2$ identity matrix *span* $\mathbb{C}^{2\times 2}$. They are noted here in the traditional form.

$$\boldsymbol{\sigma}_o = \begin{pmatrix} 1 & 0 \\ 0 & 1 \end{pmatrix}, \boldsymbol{\sigma}_1 = \begin{pmatrix} 0 & 1 \\ 1 & 0 \end{pmatrix}, \boldsymbol{\sigma}_2 = \begin{pmatrix} 0 & -i \\ i & 0 \end{pmatrix}, \boldsymbol{\sigma}_3 = \begin{pmatrix} 1 & 0 \\ 0 & -1 \end{pmatrix}. \quad (16)$$

The collection $\{\boldsymbol{\sigma}_j\}$ is the Pauli Basis, with which a transfer matrix $\boldsymbol{T}_j$ ($\boldsymbol{E}_n$ or $\boldsymbol{O}_m$) can be represented as

$$\boldsymbol{T}_j = \sum_p c_j^p \boldsymbol{\sigma}_p . \quad (17)$$

In this form $\boldsymbol{T}_j$ is identified as a *Pauli Vector*. In all the summations that follow the index runs over 0 1 2 3 unless otherwise stated. The subscript $j$ in the coefficient $c_j^p$ denotes the order of the transfer matrix while the superscript is identified with the index of the basis element it is multiplied with. Table-1 collects the coefficients for $\boldsymbol{E}_n$ and $\boldsymbol{O}_m$.

|  | $\boldsymbol{E}_n$ |  | $\boldsymbol{O}_m$ |
|---|---|---|---|
| $c_n^o$ | $\frac{\alpha}{2}(\epsilon^\beta + \epsilon^{-\beta})$ | $c_m^o$ | $\frac{\alpha}{2}(\eta^\beta + \eta^{-\beta})$ |
| $c_n^1$ | $\frac{\beta}{2}(\epsilon^\alpha + \epsilon^{-\alpha})$ | $c_m^1$ | $-\frac{\beta}{2}(\eta^\alpha + \eta^{-\alpha})$ |
| $c_n^2$ | $\frac{i\beta}{2}(\epsilon^\alpha - \epsilon^{-\alpha})$ | $c_m^2$ | $\frac{i\beta}{2}(\eta^\alpha - \eta^{-\alpha})$ |
| $c_n^3$ | $\frac{\alpha}{2}(\epsilon^\beta - \epsilon^{-\beta})$ | $c_m^3$ | $\frac{\alpha}{2}(\eta^\beta - \eta^{-\beta})$ |

**Table-1** Pauli coefficients of $\boldsymbol{E}_n$ and $\boldsymbol{O}_m$

$c_j^p$ for an arbitrary 2 × 2 matrix can also be found (equation (28)). It can be shown that the product of two transfer matrices $T_j$ and $T_k$

$$T_j T_k = \left(\sum_p c_j^p \sigma_p\right)\left(\sum_q c_k^q \sigma_q\right) = \sum_p \sigma_p \sum_q c_j^q c_k^{\phi(p,q)}(i)^{\varepsilon_{p\,q\,\phi(p,q)}}, \qquad (18)$$

$$\phi(a,b) = \left(a + b(-1)^{a+b-1}\right) \bmod 4 \qquad (19)$$

$$\varepsilon_{a\,b\,c} = \frac{1}{2}(a-b)(b-c)(c-a). \qquad (20)$$

Equation (18) results from expanding the bracketed pair and injecting the product identities of the Pauli matrices. $\varepsilon_{a\,b\,c}$ is the Levi-Civita Symbol (or Permutation symbol) which along with $\phi(a,b)$ preserves the non-commutativity of matrix multiplication. Equation (18) expresses $T_j T_k$ in the form of equation (17). This is a distinctive advantage of equation (18) *as it readily allows for matrix products to be expressed as a sum of simple matrices*. This equation is used iteratively to obtain a Pauli Vector representation for the long matrix product sequence appearing in equation (15). In the following discussion the summation indices (of equation (17)) will be augmented with an additional subscript. This will be self-explanatory to a large extent and is adopted for the sake of clarity. One begins with a single transfer matrix $T_1$ and obtains higher products as follows.

$$T_1 = \sum_{p_1} \sigma_{p_1} c_1^{p_1}$$

$$T_1 T_2 = \sum_{p_2} \sigma_{p_2} c_{12}^{p_2} = \sum_{p_2} \sigma_{p_2} \sum_{q_1} c_1^{q_1} c_2^{\phi(p_2,q_1)} (i)^{\varepsilon_{p_2\,q_1\,\phi(p_2,q_1)}}$$

$$(T_1 T_2) T_3 = \sum_{p_3} \sigma_{p_3} c_{123}^{p_3} = \sum_{p_3} \sigma_{p_3} \sum_{q_2} c_{12}^{q_2} c_3^{\phi(p_3,q_2)} (i)^{\varepsilon_{p_3\,q_2\,\phi(p_3,q_2)}}$$

$$= \sum_{p_3} \sigma_{p_3} \sum_{q_2} \sum_{q_1} c_1^{q_1} c_2^{\phi(q_2,q_1)} c_3^{\phi(p_3,q_2)} (i)^{\varepsilon_{q_2\,q_1\,\phi(q_2,q_1)} + \varepsilon_{p_3\,q_2\,\phi(p_3,q_2)}}$$

$$(T_1 T_2 T_3) T_4 = \sum_{p_4} \sigma_{p_4} c_{1234}^{p_4} = \sum_{p_4} \sigma_{p_4} \sum_{q_3} c_{123}^{q_3} c_4^{\phi(p_4,q_3)} (i)^{\varepsilon_{p_4\,q_3\,\phi(p_4,q_3)}}$$

$$= \sum_{p_4} \sigma_{p_4} \sum_{q_3} \sum_{q_2} \sum_{q_1} c_1^{q_1} c_2^{\phi(q_2,q_1)} c_3^{\phi(q_3,q_2)} c_4^{\phi(p_4,q_3)} (i)^{\varepsilon_{q_2\,q_1\,\phi(q_2,q_1)} + \varepsilon_{q_3\,q_2\,\phi(q_3,q_2)} + \varepsilon_{p_4\,q_3\,\phi(p_4,q_3)}}$$

setting $q_0 = 0$ and noting that $q_1 = \phi(q_1, 0) = \phi(q_1, q_0)$ the general formula can be written as,

$$T_1 T_2 \ldots T_{M-1} T_M$$
$$= \sum_{p_M} \sigma_M \sum_{q_{M-1}} \sum_{q_{M-2}} \ldots \sum_{q_3} \sum_{q_2} \sum_{q_1} \prod_{j=1}^{M-1} \left( c_j^{\phi(q_j,q_{j-1})} (i)^{\varepsilon_{q_j\,q_{j-1}\,\phi(q_j,q_{j-1})}} \right) c_M^{\phi(p_M,q_{M-1})} (i)^{\varepsilon_{p_M\,q_{M-1}\,\phi(p_M,q_{M-1})}}.$$

But for the outer most summation, the inner multiple summation is a *scalar*. This inductive construction will be of great use in section V when the product of $M$ transfer matrices is need for the generalized NBP. For the problem at hand, equation (15) can be mapped as $T_1 \to E_o$, $T_2 \to O_1$, $T_3 \to E_2 \ldots T_M \to O_{2N-1}$ to obtain the required representation. To avoid the long formula this product is denoted as

$$E_0 O_1 E_2 \ldots E_{2N-2} O_{2N-1} = \sum_p c_{NBP}^p \sigma_p, \qquad (21)$$

where the $c_{NBP}^p$s can be readily obtained using the above prescription.

The standard orthogonal vectors listed in equation (22) display interesting algebraic relations upon multiplication by the Pauli Matrices.[16]

$$|+\rangle = \begin{pmatrix} 1 \\ 0 \end{pmatrix} \qquad |-\rangle = \begin{pmatrix} 0 \\ 1 \end{pmatrix} \quad . \tag{22}$$

From equations (14) and (22) it follows

$$T = \frac{\langle V_{2N}|+\rangle\langle +|V_{2N}\rangle}{\langle V_o|+\rangle\langle +|V_o\rangle} \qquad R = \frac{\langle V_o|-\rangle\langle -|V_o\rangle}{\langle V_o|+\rangle\langle +|V_o\rangle} \quad . \tag{23}$$

$\langle +|V_o\rangle$ and $\langle -|V_o\rangle$ can be calculated using equation (15)

$$\langle \pm|V_o\rangle = \frac{\langle \pm|E_0 O_1 E_2 \ldots E_{2N-2} O_{2N-1}|V_{2N}\rangle}{\Delta^N}$$

$$= \frac{A_{2N}}{\Delta^N} \langle \pm| \sum_p \sigma_p c^p_{NBP} |+\rangle$$

$$= \frac{A_{2N}}{\Delta^N} \sum_p c^p_{NBP} \langle \pm|\sigma_p|+\rangle$$

$$\Rightarrow \begin{cases} \langle +|V_o\rangle = \frac{A_{2N}}{\Delta^N}(c^0_{NBP} + c^3_{NBP}) \\ \langle -|V_o\rangle = \frac{A_{2N}}{\Delta^N}(c^1_{NBP} + i c^2_{NBP}) \end{cases} \quad . \tag{24}$$

Substituting equation (24) and $\langle +|V_{2N}\rangle = A_{2N}$ in equation (23)

$$T = \left| \frac{\Delta^N}{c^0_{NBP} + c^3_{NBP}} \right|^2 \qquad R = \left| \frac{c^1_{NBP} + i c^2_{NBP}}{c^0_{NBP} + c^3_{NBP}} \right|^2 \tag{25}$$

Equation (25) collects the expressions for the transmission and reflection coefficients. Note that these are *independent* of $A_{2N}$. For $E = V_o$, $\Delta$ goes to zero, hence $T$ and $R$ as given in (25) are indeterminate (Ref. Equation (15)). This ambiguity can be resolved by explicitly solving the TISE for $E = V_o$. In that case equation (2) leads to solutions of the form $\psi_m = A_m x + B_m$. Using these in equation (9) gives the correct transfer matrices and the subsequent procedure is exactly identical.

**IV DISCUSSION OF RESULTS**

$T$ turns out to be exceedingly complex for large barrier numbers. Thus it can best be expressed in an *implicit* form. To discuss effectively about the nature of $T$(as a function of $E$), $\kappa$ is identified as a natural variable in tunneling problems. In the units of $\frac{2\mu}{\hbar^2}$, $\kappa$ equals $\sqrt{E}$. Closed form expressions for $T$ and $R$ for a rectangular barrier and potential step (no barrier) are well known.[14] These trivial cases are illustrated using this method.

The potential step as per the indexing convention(Ref. Fig.1) has only two regions(0 and $I$). Thus $|V_1\rangle = \begin{pmatrix} A_1 \\ o \end{pmatrix} = A_1|+\rangle$.

$$|V_o\rangle = \frac{1}{2\gamma} E_o |V_1\rangle = \frac{A_1}{2\gamma}(\alpha \sigma_o + \beta \sigma_1)|+\rangle = \frac{A_1}{2\gamma}\begin{pmatrix} \kappa + \gamma \\ \kappa - \gamma \end{pmatrix} = \begin{pmatrix} A_o \\ B_o \end{pmatrix}$$

For the case of no barriers, $T$ and $R$ as defined in [14] are obtained as.

$$R = \left| \frac{\kappa - \gamma}{\kappa + \gamma} \right|^2 \qquad T = \frac{4\kappa\gamma}{|\kappa + \gamma|^2}$$

As a consistency check, $T + R = 1$. For a rectangular barrier correspondingly, there are three regions and $|V_2\rangle = \begin{pmatrix} A_2 \\ o \end{pmatrix} = A_2|+\rangle$. From equation 3.15

$$|V_o\rangle = \frac{1}{2\gamma} E_o |V_1\rangle = \frac{1}{\Delta} E_o O_1 |V_2\rangle$$

$$= \frac{A_2}{\Delta}(\alpha \sigma_o + \beta \sigma_1)(c^0_1 \sigma_o + c^1_1 \sigma_1 + c^2_1 \sigma_2 + c^3_1 \sigma_3) |+\rangle$$

$$= \frac{A_2}{\Delta}\begin{pmatrix} \alpha c_1^o + \alpha c_1^3 + \beta c_1^1 + i\beta c_1^2 \\ \alpha c_1^1 + i\alpha c_1^2 + \beta c_1^o + \beta c_1^3 \end{pmatrix},$$

$\eta = e^{i\delta}$ for $m = 1$ and from Table-1, $c_1^o = \alpha \cos(\beta\delta)$, $c_1^1 = -\beta \cos(\alpha\delta)$, $c_1^2 = -\beta \sin(\alpha\delta)$, $c_1^3 = i\alpha \sin(\beta\delta)$. Substituting these coefficients gives

$$|V_o\rangle = \frac{A_2}{\Delta}\begin{pmatrix} \alpha^2 e^{i\beta\delta} - \beta^2 e^{i\alpha\delta} \\ \alpha\beta(e^{i\beta\delta} - e^{i\alpha\delta}) \end{pmatrix} = \begin{pmatrix} A_o \\ B_o \end{pmatrix}.$$

From equation (23)

$$R = \left|\frac{\alpha\beta(e^{i\beta\delta} - e^{i\alpha\delta})}{\alpha^2 e^{i\beta\delta} - \beta^2 e^{i\alpha\delta}}\right|^2 \qquad T = \left|\frac{\Delta}{\alpha^2 e^{i\beta\delta} - \beta^2 e^{i\alpha\delta}}\right|^2.$$

These expressions are arrived at more economically than with the standard approach.

Equation (25) leads to much faster computation and plots for very large number of barriers can be easily obtained. The trends that set in, as the barrier number is progressively increased are discussed next. All the plots were obtained using MATLAB® 7.10.0.499. In Fig.2 several plots for NBPs over an energy range of 81 units are given. They correspond to increasing values of $N$. The specifications are $\delta = \tau = 1$ and $V_o = 40$. These values were chosen for comparing with the results given by Rao et al.[13] In Fig.2a the case for $N = 1$ and $N = 2$ is plotted using equation (25) and an *exact* reproduction of Figure2 of their paper is obtained. The value of $N$ is given in the figures.

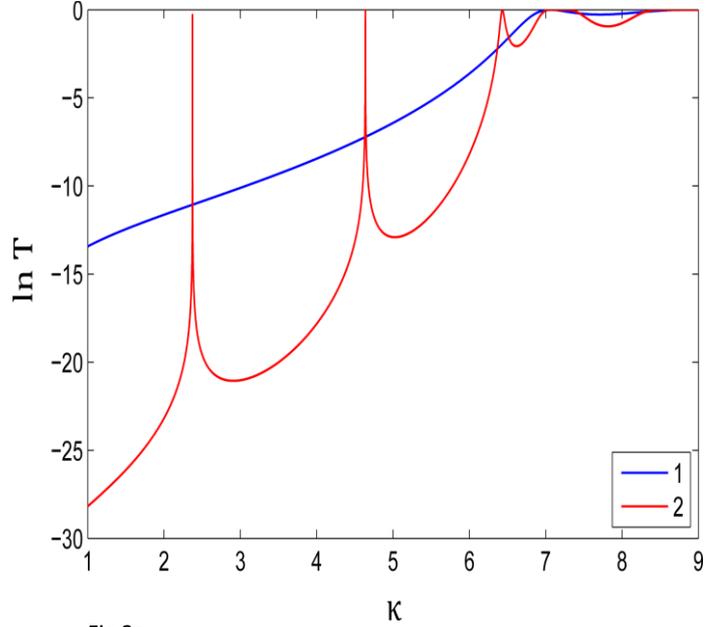
Fig.2a

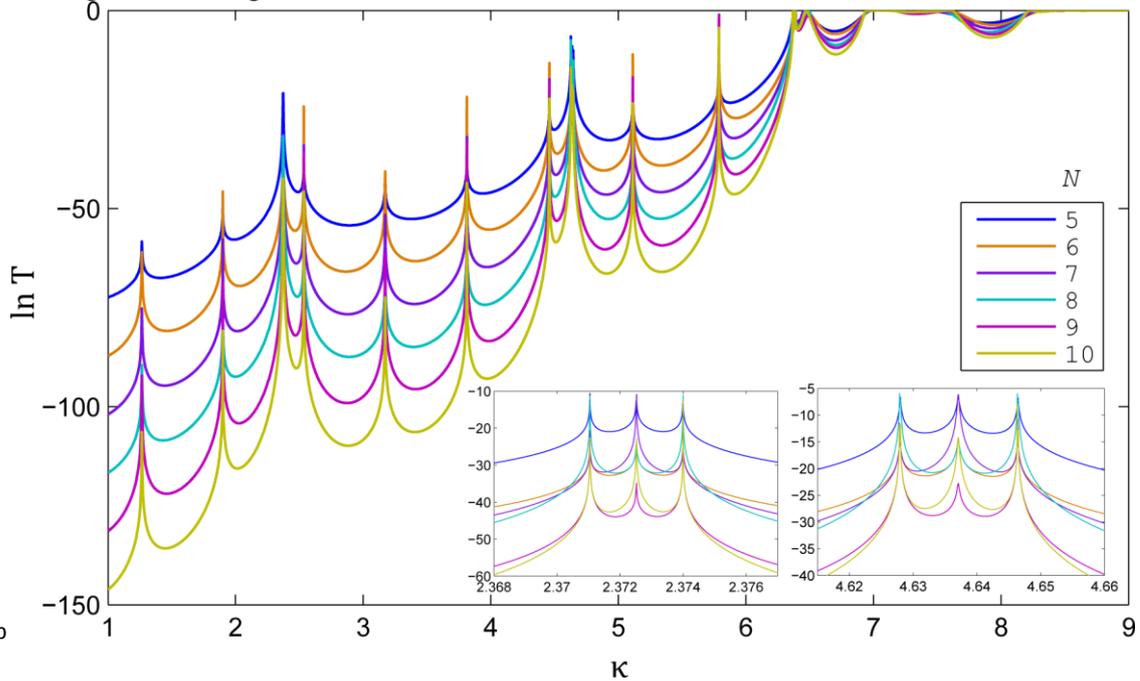
Fig.2b

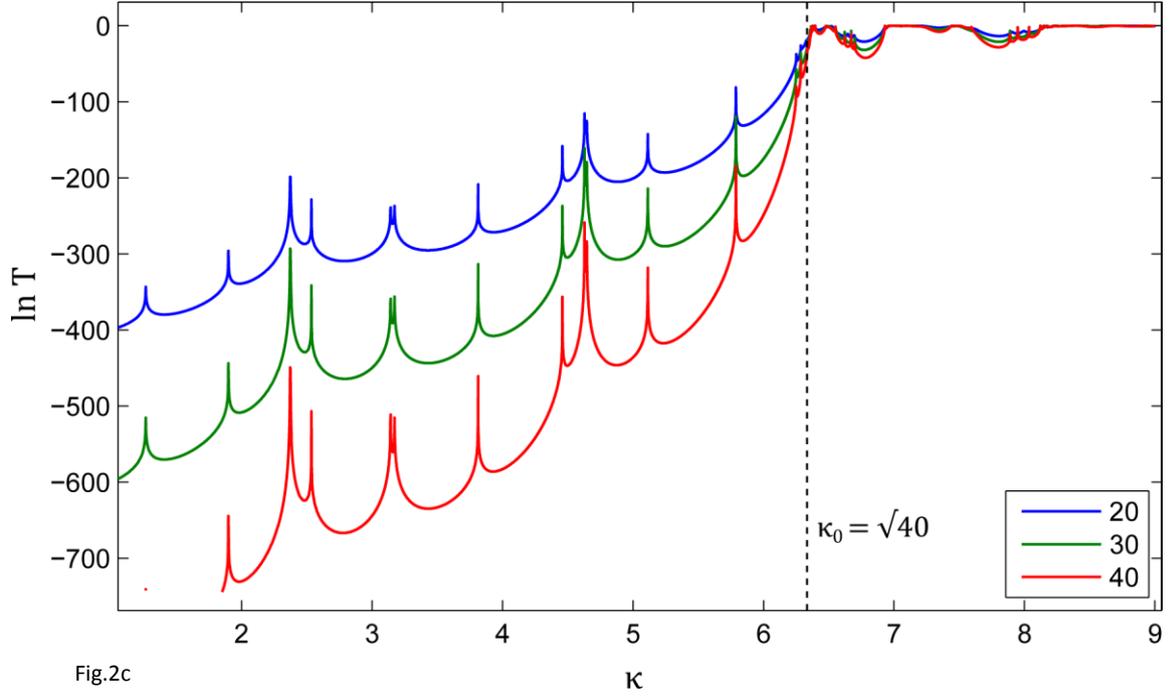

Fig.2c

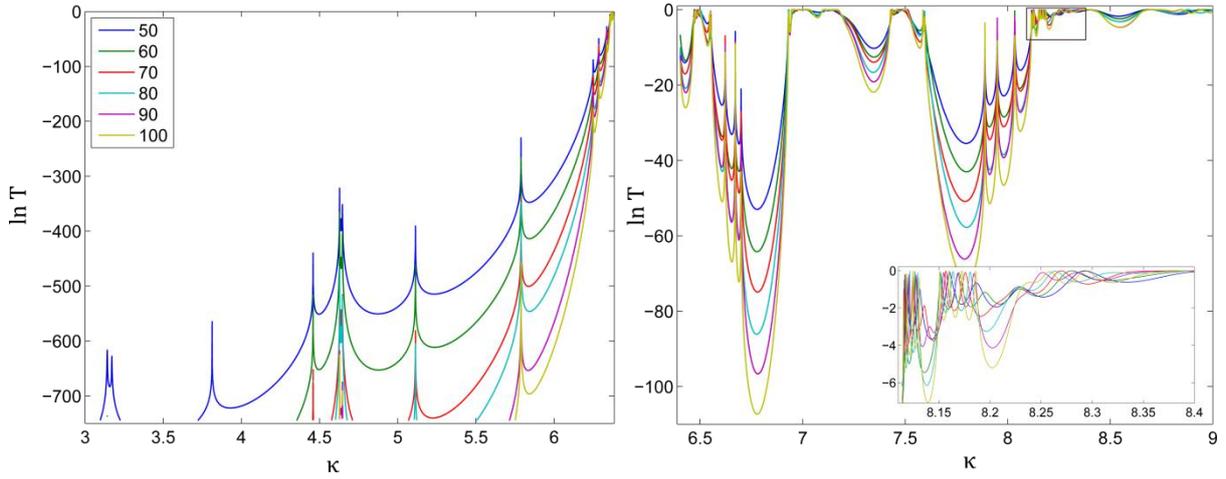

Fig.2d

**Fig 2:** $\ln T$ vs $\kappa$ for barrier parameters : $\tau = \delta = 1$, $V_o = 40$. $\kappa_o = 6.32$. $N$ denotes number of barriers.

a. $N = 1$ and $N = 2$
b. $N = 3$, $N = 4$, $N = 5$, $N = 6$, $N = 7$, $N = 8$, $N = 9$, $N = 10$. The insets depict the splitting of spikes at $\kappa = 4.6$ and $\kappa = 2.3$
c. $N = 20$, $N = 30$, $N = 40$. $\kappa_o$ is marked by broken vertical line.
d. $N = 50$, $N = 60$, $N = 70$, $N = 80$, $N = 90$, $N = 100$. Inset magnifies the region contained in the rectangle.

The transmission characteristics of a NBP can be analyzed by contrasting it with the classical solution where $T$ vs. $\kappa$ takes the form of a step function :

$$T = \begin{cases} 0 & \kappa < \sqrt{V_o} \\ 1 & \kappa \geq \sqrt{V_o} \end{cases} . \tag{26}$$

The threshold value $\sqrt{V_o}$ is denoted as $\kappa_o$. For the given plots $\kappa_o = \sqrt{40} = 6.32$. Clearly the transmission probability $T$ decreases for $\kappa < \kappa_o$ with increasing $N$, however asymptotes closer to 1 (always staying below unity) for $\kappa \geq \kappa_o$. In fact when $N$ is large enough such that the overall width of the barrier: $N\delta + (N-1)\tau$ is comparable to (say) an average *classical length*, it is conceivable that $T$ approaches the classical characteristics of (26).

The features that are striking are the resonant spikes for $\kappa < \kappa_o$. These spikes generally correspond to the bound states of the interlocked wells: $\kappa = \frac{l}{\tau}\pi$, (for $\kappa \ll \kappa_o$) while this is not strictly obeyed(especially for large $l$ or higher bound energies) due to coupling effects; many extraneous spikes also appear. All spikes naturally extend to unity. For $\kappa < \kappa_o$ the resonant spikes are relatively insensitive to variations in barrier width $\delta$ and stay fixed for a given choice of barrier parameters for different values of $N$. All the curves predominantly stay disjoint in the region $\kappa < \kappa_o$ except at the resonant points, where spikes arise and touch unity, while the curves cross each other at many points for $\kappa \geq \kappa_o$ and this region is marked by small oscillatory excursions below unity(Fig.2c,Fig.2d). Some spikes *split* into two or three individual spikes. These are so narrow that one faces the problem of resolving them, given the constraints of under-sampling by the plotting software. Two triple spikes are shown as insets in Fig.2b. The transmission characteristics for 500 and 1000 barriers are depicted in Fig.3. The barrier specifications are same as those of Fig.2. One can observe a behavior of varying complexity with smooth depressions of $T \approx e^{-600}$ for $N = 1000$ (not shown in figure) and regions with resonant spikes that occur in groups. Thus resonant energy states group into bands separated by forbidden gaps (wells).

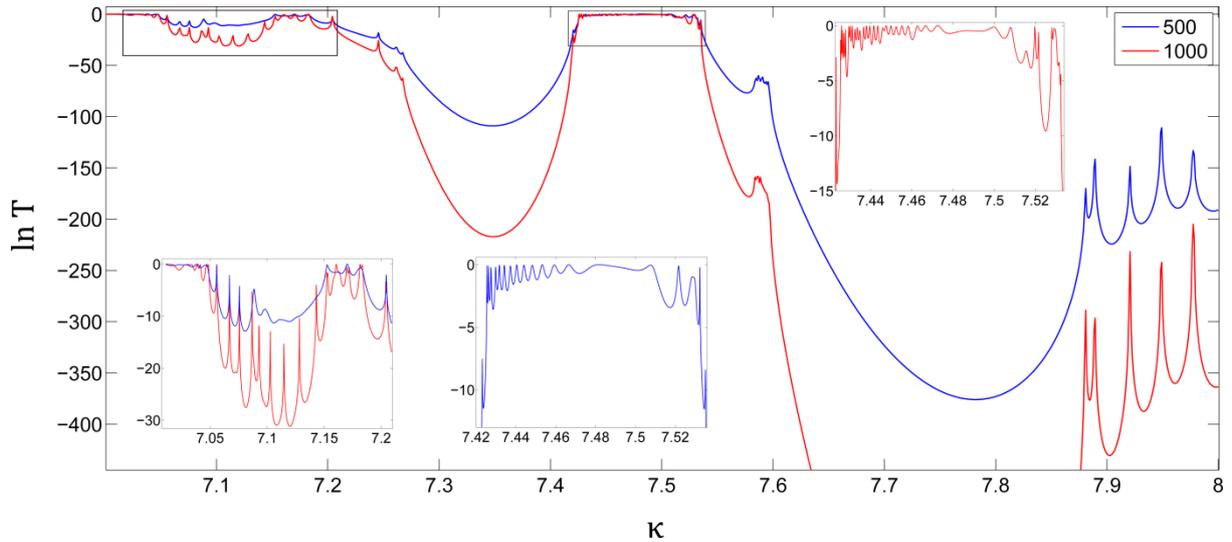

**Fig 3:** $\ln T$ vs $\kappa$ for 500 and 1000 barriers for $\tau = \delta = 1, V_o = 40$. Insets depict details of boxed regions.

Keeping $N$ fixed if one increases $V_o$, the curves again approach the classical characteristics, which is reconcilable as the barrier height progressively takes values of significance in classical energy scales- even then the quantum corrections do not disappear altogether and persistent depressions develop for $\kappa \geq \kappa_o$. Fig.4 illustrates this for the case of 9 barriers (Fig.4a) and 30 barriers (Fig.4b) at decade changes in $V_o$. Corresponding $\kappa_o$s are also shown in the figure by means of broken vertical lines.

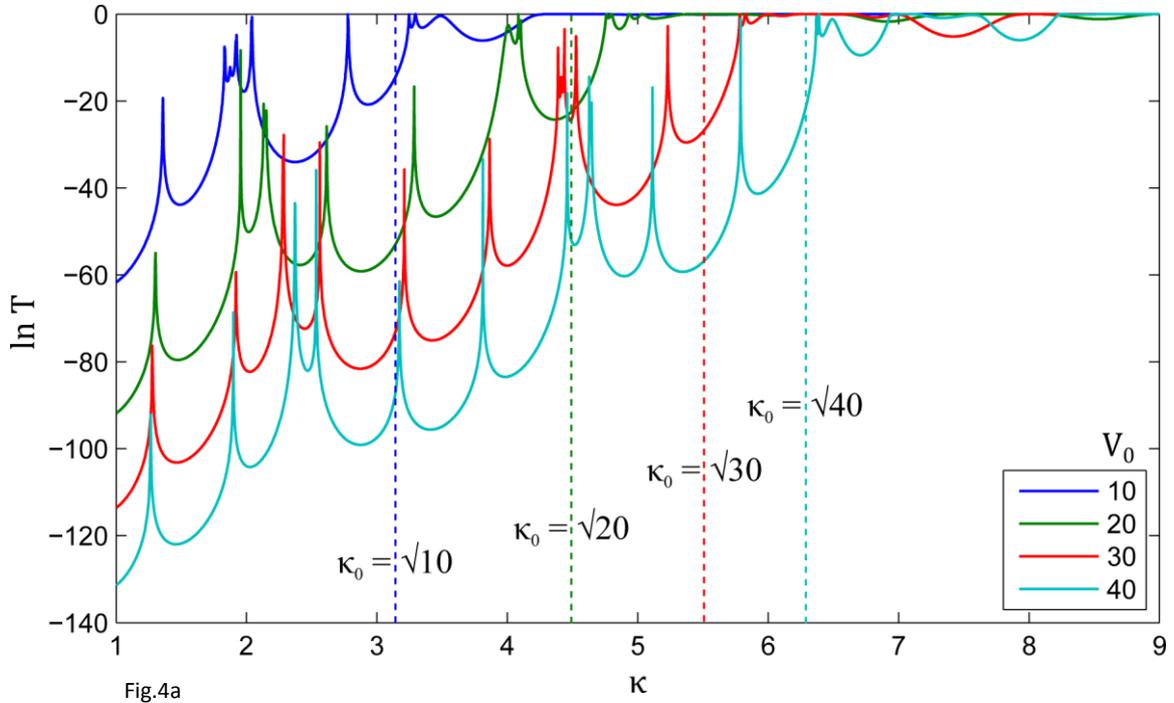

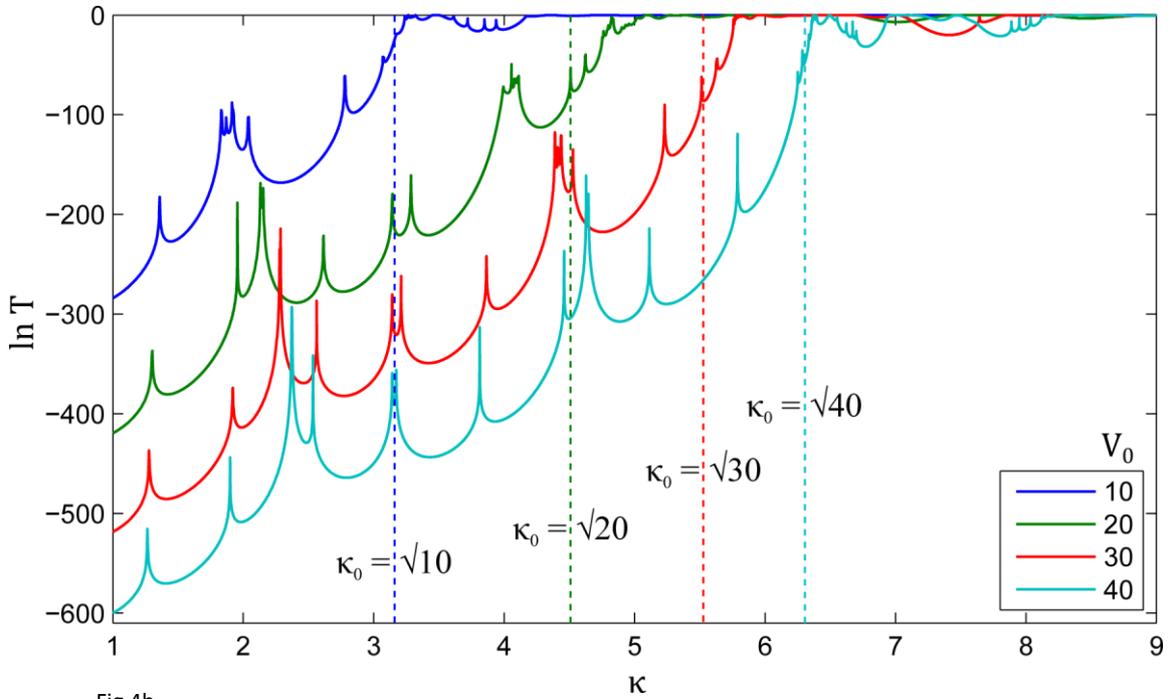

**Fig 4:** $\ln T$ vs $\kappa$. for barrier parameters: $\tau = \delta = 1$. The different curves are at decade separation of $V_o$. These are noted in the legend and the corresponding $\kappa_o$s are depicted by means of broken vertical lines. Note that the resonant spikes shift right with increasing $V_o$.

a. All curves are for 9 barriers

b. All curves are for 30 barriers

An interesting feature that manifests for large $N$ is the appearance of predominant *'probability wells'* in the region $\kappa \geq \kappa_o$. Some of the oscillations of $T$ develop into predominant depressions, flattening and deepening(towards zero) with increasing $N$. Approaching zero these 'flat' regions attain a finite width. Until this point the walls of these 'probability wells' remain smooth and unperturbed. Further increase in barrier number gives rise to spikes along the walls(Fig.5). These spikes develop by the impregnation of shallower local minimas at the brim of a well that narrow down and deepen, rendering their interspersed maxima into very thin spikes. The situation is fairly symmetric on either side of a well. The spikes, lining a well further grow in number with increasing barrier number ($N = 300$, $N = 400$, for instance). And the wells shift position with changing barrier height $V_o$. To be able to appreciate this effect it is instructive to plot $T$ (not $\ln T$) against $\kappa$-as the spikes become more predominant. However this gives the impression that the wells touch zero sufficiently fast in $N$, due to the rounding off schemes employed in the plotting software. So the convention of plotting $\ln T$ is retained in the accompanying figures.

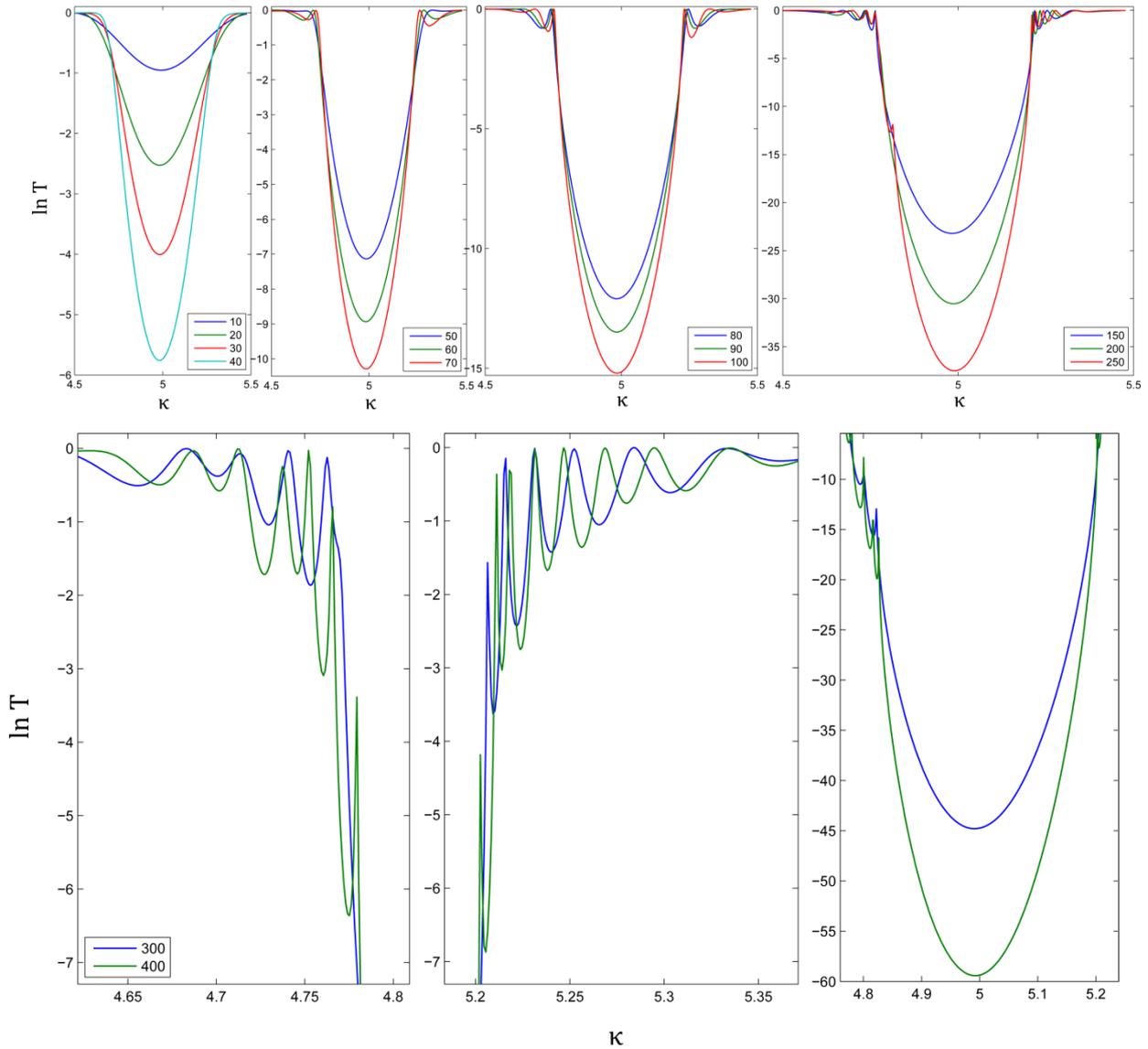

**Fig 5**: Probability wells for $\tau = \delta = 1$, $V_o = 5$. Number of barriers $N$ is noted in the legend. These are forbidden bands of energy where the tunneling probability falls strongly with increasing barrier number.

These plots open up a whole new set of possibilities both for the theoretician as well as for the application minded researcher. Given the prowess of Integrated Circuit fabrication techniques, realization of barrier structures with suitable choice of parameters is not difficult to imagine. This presentation may trigger novel innovations in electronic technology and many other areas. The analytic investigation of these curves using the given formula amounts to a treatise on its own as interesting patterns emerge upon deeper investigation (although it poses a formidable mathematical challenge). These may be taken up in a future paper. Hence further observations are not provided. The main aim of this work was to provide the solution methodology and point out some immediate consequences for large $N$.

## V A NOTE ON GENERALIZATION AND CONCLUSION:

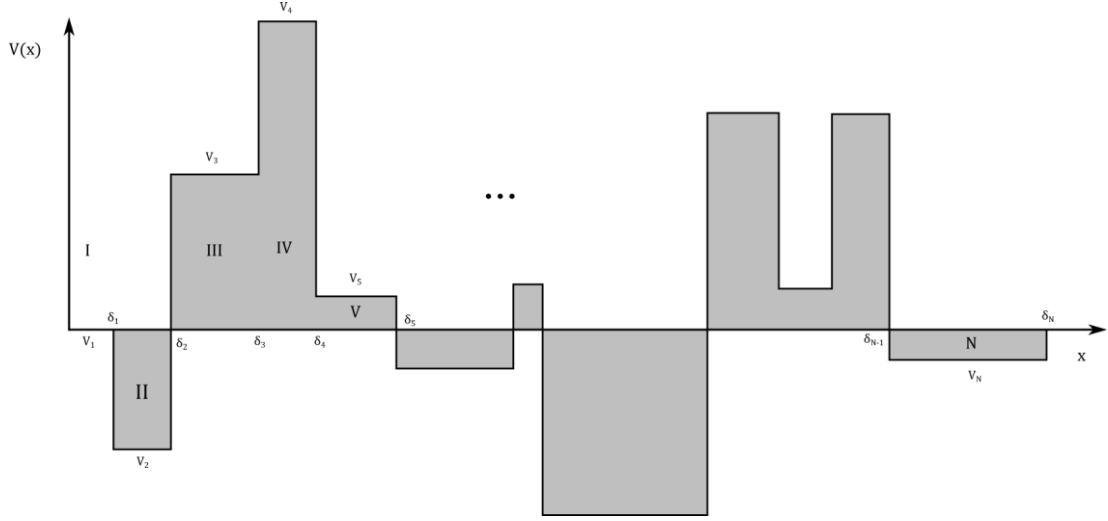

**Fig.6:** $V(x)$ for an arbitrary piecewise constant N Barrier Problem.

This section is presented for the sake of completeness. Here the problem is extended to a general piece wise-constant potential barrier. A generic NBP can be defined by specifying two sequences, $\{V_j\}$ and $\{\delta_j\}$. $\{\delta_j\}$ must be necessarily positive and monotonically increasing. These sequences derive their meaning from Fig.6. Most of the considerations of Section III hold good. In this case a single continuity condition suffices that can be applied for a general point $\delta_j$. And similar transfer matrix equations are obtained:

$$|V_j\rangle = \frac{1}{2\kappa_j} \boldsymbol{T}_j |V_{j+1}\rangle , \qquad (27)$$

where

$$\boldsymbol{T}_j = \begin{pmatrix} \alpha_j \epsilon^{\beta_j} & \beta_j \epsilon^{\alpha_j} \\ \beta_j \epsilon^{-\alpha_j} & \alpha_j \epsilon^{-\beta_j} \end{pmatrix}$$

$$\alpha_j = \kappa_j + \kappa_{j+1} \qquad \beta_j = \kappa_j - \kappa_{j+1} \qquad \kappa_l^2 = \frac{2\mu}{\hbar^2}(E - V_l) \qquad \epsilon = e^{-i\delta_j}$$

Now $\boldsymbol{T}_j$ can be written in the form of equation (17) where the generalized coefficients $c_j^p$ are given by

$$c_j^p = \text{Trace}(\boldsymbol{\sigma}_p \boldsymbol{T}_j). \qquad (28)$$

The remaining computations are analogous. The derivation for the general transfer matrix product was given in Section III.

From this generalization an approximation scheme for estimating $T$ and $R$ for more realistic barrier potentials that are continuous(or smooth) can be formulated. The given $V(x)$ can be sampled at many points by taking a partition on $x$. Thus obtaining a step approximation to the potential and reducing the problem to that of a piecewise constant potential NBP. Following which the general approach can be used to calculate the transmission coefficient. The accuracy can be made arbitrarily good by refining the partition on $x$.

A possible extension would be for multidimensional tunneling, i.e. one in which a potential field, like $V(x,y)$ is considered. Again if $V$ is well behaved in the region of interest, the use of variable separable method for solving the TISE will lead to similar one dimensional NBPs for every direction and the methods described here can be adopted. Though in this case $T$ has to be defined more precisely. A consistent discussion of one dimensional tunneling across $N$ rectangular Barriers is provided in this paper. Some of the topics that have not been touched are tunneling time and tunneling length. These are interesting parameters to look at for a NBP. Time evolution of the wave function is another aspect that requires further insight. Certainly the analysis of these problems rests directly on the discussion provided in this paper. Also, long sequence matrix products of the form presented here, call for optimal computational algorithms that reduce code and time complexities.

The author extends his acknowledgement to Dr.Pankaj Agarawal (Institute of Physics, Bhubaneswar), Dr.Hemalatha Thiagarajan (NIT,Trichy) and Dr.T.N.Janakiraman (NIT,Trichy) for the many ways in which they have contributed to the completion of this project. Sambhav R Jain (Design Engineer, Texas Instruments) has prepared the excellent graphics for this paper in Inkscape.

Siddhant Das
8|10|2013